\documentclass[preprint,showpacs,preprintnumbers,amsmath,amssymb,aps]{revtex4}

\usepackage{graphicx}
\usepackage{dcolumn}
\usepackage{bm}

\begin{document}

\preprint{APS/123-QED}

\title{Anomalous magnetization of $S$=2 ferromagnetic-antiferromagnetic alternating chain with Ising anisotropy}

\author{H. Yamaguchi$^1$, Y. Shinpuku$^1$, Y. Kono$^2$, S. Kittaka$^2$, T. Sakakibara$^2$, M. Hagiwara$^3$, T. Kawakami$^4$, K. Iwase$^1$, T. Ono$^1$, and Y. Hosokoshi$^1$}
\affiliation{
$^1$Department of Physical Science, Osaka Prefecture University, Osaka 599-8531, Japan \\ 
$^2$Institute for Solid State Physics, the University of Tokyo, Chiba 277-8581, Japan \\
$^3$Center for Advanced High Magnetic Field Science (AHMF), Graduate School of Science, Osaka University, Osaka 560-0043, Japan \\
$^4$Department of Chemistry, Osaka University, Osaka 560-0043, Japan \\
}

Second institution and/or address\\
This line break forced

\date{\today}

\begin{abstract}
We present the first experimental realization of an $S=2$ ferromagnetic-antiferromagnetic (F-AF) alternating chain in a new  Mn-verdazyl complex [Mn(hfac)$_2$]$\cdot$($o$-Py-V) [hfac=1,1,1,5,5,5-hexafluoroacetylacetonate; $o$-Py-V=3-(2-pyridyl)-1,5-diphenylverdazyl].
Through the $ab$ $initio$ molecular orbital calculation, magnetization, and ESR measurements, this compound is confirmed to form an $S=2$ F-AF alternating chain with Ising anisotropy below about 100 K. 
Furthermore, we find an anomalous change in magnetization at 1/4 of the saturation value, which is probably a manifestation of the quantum nature of the system.
 \end{abstract}

\maketitle
Quantum spin systems provide a variety of unique many-body phenomena through strong quantum fluctuations.
Haldane's prediction in 1983~\cite{Haldane} has stimulated experimental and theoretical studies focused on the qualitative difference between half-integer and integer spin Heisenberg antiferromagnetic (AF) chains.
Subsequent investigations on $S$=1 systems have established the presence of a Haldane gap in the integer spin case~\cite{Todo, Nakano, NENP, CsNiCl3}.
The ground state below the Haldane gap is well described as a valence-bond-solid (VBS)~\cite{Affleck, Hagiwara}, where each integer spin is considered as two half-size spins forming a singlet state between different sites.
However, there are remarkably few examples of $S{\textgreater}$1 quantum spin systems. 
The second-minimal integer spin, $S$=2, has only several examples~\cite{CsCrCl3, MnCl3(bipy), a-NaMnO2, MnF(salen), CrCl2, CsCrCl3_2}, and their quantum behavior still has not been clarified.
The spin degrees of freedom for $S$=2 systems can be expected to create rich quantum states.
The ground-state phase diagram on the $S$=2 AF chain is discussed in several numerical works~\cite{S2_phase1,S2_phase2,S2_phase3,S2_phase4,S2_phase5,S2_phase6,S2_phase7,S2_phase8}, where varieties of quantum phase appear in accordance with the combination of exchange and on-site anisotropies.
Moreover, a bond alternation in the $S$=2 AF chain is expected to induce a hidden symmetry breaking in the VBS state~\cite{S2_alt1,S2_alt2,S2_alt3}.

The VBS of the integer spin chain can be mapped onto a strong ferromagnetic (F) coupling limit of an F-AF alternating chain with half-size spin.
Therefore, the $S$=1/2 Heisenberg F-AF chain has been intensively studied in relation with the Haldane state in the $S$=1 Heisenberg AF chain~\cite{sakai,neut1, neut2}. 
Hida investigated the ground-state properties for various ratios of exchange constants~\cite{hida1, hida2, hida3}.
The energy gap and string order parameter, which indicates hidden topological order specific to VBS, have finite values for all the ratios of exchange constants and change continuously from the $S$=1 Haldane ($|J_{\rm{F}}| \gg J_{\rm{AF}}$) phase to the $S$ = 1/2 AF dimer ($|J_{\rm{F}}| \ll J_{\rm{AF}}$) phase.
Thus, it is suggested that there is no discontinuous change in the ground state associated with a phase transition between two opposite limiting cases. 
Accordingly, the ground-state properties of the present $S$=2 F-AF alternating chain could be related to that of an $S$ = 4 AF chain, even though the details of exchange path slightly differs from that of the $S$=1/2 case.  

Recently, we have established synthetic techniques for preparing high-quality verdazyl-based single crystals, and we have demonstrated modulations of the magnetic interactions by using simple chemical modifications~\cite{3Cl4FV, mPhV2, 26Cl2V, fine-tune}.
More recently, we have successfully proceeded to the next stage of the synthesis, wherein verdazyl radical coordinates to transition metal~\cite{Zn}.
Such complexes enable us to manipulate the spin size through the strong coupling between radical and magnetic ion; this technique is the focus of the present work.

In this letter, we report the first experimental realization of an $S$=2 F-AF alternating chain with Ising anisotropy.
We have succeeded in synthesizing a new Mn-verdazyl complex [Mn(hfac)$_2$]$\cdot$($o$-Py-V), wherein hfac is equal to 1,1,1,5,5,5-hexafluoroacetylacetonate, and $o$-Py-V is equal to 3-(2-pyridyl)-1,5-diphenylverdazyl.
The $ab$ $initio$ molecular orbital (MO) calculation indicates the formation of the $S$ = 2 F-AF alternating chain.
The magnetic properties show remarkably anisotropic behavior, such as metamagnetic phase transition. 
We evaluated on-site anisotropy from the ESR analysis.  
Furthermore, we observed an anomalous change in magnetization at 1/4 of the saturation value.

The synthesis of [Mn(hfac)$_2$]$\cdot$($o$-Py-V), whose molecular structure is shown in Fig. 1(a), was performed using a procedure similar to that for [Zn(hfac)$_2$]$\cdot$($o$-Py-V)~\cite{Zn}.
The crystal structure was determined on the basis of intensity data collected using a Rigaku AFC-8R Mercury CCD RA-Micro7 diffractometer with Japan Thermal Engineering XR-HR10K. 
The magnetizations were measured using a commercial SQUID magnetometer (MPMS-XL, Quantum Design) and a capacitive Faraday magnetometer down to about 70 mK.
The experimental results were corrected for diamagnetic contribution calculated by Pascal's method.
The specific heat was measured by a standard adiabatic heat-pulse method down to about 0.1 K. 
The ESR measurements were performed utilizaing a vector network analyzer (ABmm), a superconducting magnet (Oxford Instruments), and a home-made ESR cryostat at AHMF in Osaka University.

The crystal structure at room temperature (RT) is isomorphous to that of [Zn(hfac)$_2$]$\cdot$($o$-Py-V)~\cite{Zn} (see Table S1).
We focus on the low-temperature structure at 25 K in order to consider magnetic interactions that act in a dominant fashion below about 100 K.
Although the space group slightly changes from $P2_{1}/c$ at RT to $P2_{1}$ at 25 K, there is little change in the molecular arrangements (see Supplementary). 
The crystallographic parameters at 25 K are as follows~\cite{CCDC}: monoclinic, space group $P2_{1}$, $a$=8.7586(12) $\rm{\AA}$, $b$=31.988(5) $\rm{\AA}$, $c$=10.8617(15) $\rm{\AA}$, $\beta$=92.568(4)$^{\circ}$, and $V$=3040.1(7) $\rm{\AA}^3$.
We performed MO calculations in order to evaluate intramolecular interaction between spins on Mn$^{\rm{2+}}$, $S_{\rm{Mn}}$ = 5/2, and verdazyl radical, $S_{\rm{V}}$=1/2, and the intermolecular interactions between their resultant spins ~\cite{MOmetho}.
The intramolecular interaction $J_{\rm{intra}}$, which is defined by the spin Hamiltonian as $\mathcal {H}=J_{\rm{intra}}S_{\rm{Mn}}{\cdot}S_{\rm{V}}$, is evaluated to be about 664 K.
This strong AF $J_{\rm{intra}}$ gives rise to a large energy gap of 3$J_{\rm{intra}}$ between resultant spins of $S$=2 and 3, as shown in Fig. 1(b).
Because the lower states of $S$=2 are well separated from the excited states of $S$=3, the [Mn(hfac)$_2$]$\cdot$($o$-Py-V) molecule can be considered to have $S$ = 2 for $T \ll 3J_{\rm{intra}}$.
The MO calculation indicated that approximately 70 \% of the total spin density of $S$ = 2 is present on the Mn atom, whereas the $o$-Py-V has approximately 28 \% of the total spin density of $S$ = 2.
The hfac works as a spacer between molecular spins, resulting in the low dimensionality of the magnetic lattice

Furthermore, intermolecular couplings arise only through the overlapping of the molecular orbitals on the $o$-Py-V.   
We found two types of dominate interactions through the $ab$ $initio$ MO calculations.
A strong F interaction $J_{\rm{F}}$ is evaluated between molecules labeled as M$_0$ and M$_1$, as shown in Fig. 1(c).
The M$_0$-M$_1$ molecular pair has two N-C short contacts $d_1$ and $d'_{1}$, which are approximately 3.35 and 3.42 $\rm{\AA}$, respectively.
An AF interaction $J_{\rm{AF}}$ is evaluated between molecules labeled as M$_0$ and M$_2$, as shown in Fig. 1(c).
The M$_0$-M$_2$ molecular pair has two C-C short contacts $d_2$ and $d'_{2}$, which are approximately 3.31 and 3.33 $\rm{\AA}$, respectively.
Those two molecular pairs are alternately aligned along the $c$-axis, as shown in Fig. 1(c), and form an $S$ = 2 F-AF alternating chain consisting of $J_{\rm{F}}$ and $J_{\rm{AF}}$ in analogy with that for [Zn(hfac)$_2$]$\cdot$($o$-Py-V)~\cite{Zn}, as shown in Fig. 1(d). 
The value of the interactions are evaluated as $J_{\rm{F}}/k_{\rm{B}}$ = $-$1.21 K and $J_{\rm{AF}}/k_{\rm{B}}$ = 0.26 K ($|J_{\rm{AF}}/J_{\rm{F}}|$=0.21), which are defined in the spin Hamiltonian given by
\begin{equation}
\mathcal {H} = J_{\rm{F}}{\sum^{}_{i}}\textbf{{\textit S}}_{2i}{\cdot}\textbf{{\textit S}}_{2i+1}+J_{\rm{AF}}{\sum^{}_{i}}\textbf{{\textit S}}_{i}{\cdot}\textbf{{\textit S}}_{2i-1}+D{\sum^{}_{i}}(\mbox{$S$}^{z}_{i})^2,
\end{equation}
where $D$ is an on-site anisotropy, the $z$-axis corresponds to the $b$-axis (discussed later), and $\textbf{{\textit S}}$ is an $S$=2 spin operator.

\begin{figure}[t]
\begin{center}
\includegraphics[width=18pc]{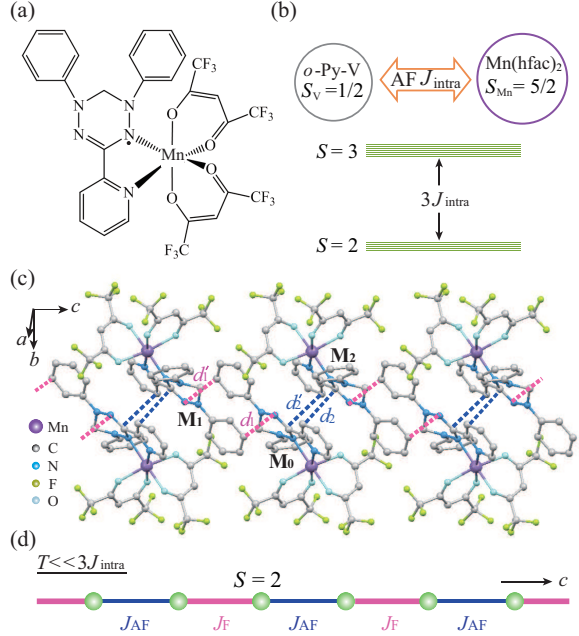}
\caption{(color online) (a) Molecular structure of [Mn(hfac)$_2$]$\cdot$($o$-Py-V). (b) Schematic view of intramolecular interaction between spins on Mn$^{\rm{2+}}$ and verdazyl radical and resulting spin states. (c) Crystal structure forming an alternating chain along the $c$-axis, and (d) the corresponding spin model.  Hydrogen atoms are omitted for clarity.}\label{f1}
\end{center}
\end{figure}

\begin{figure}[t]
\begin{center}
\includegraphics[width=16pc]{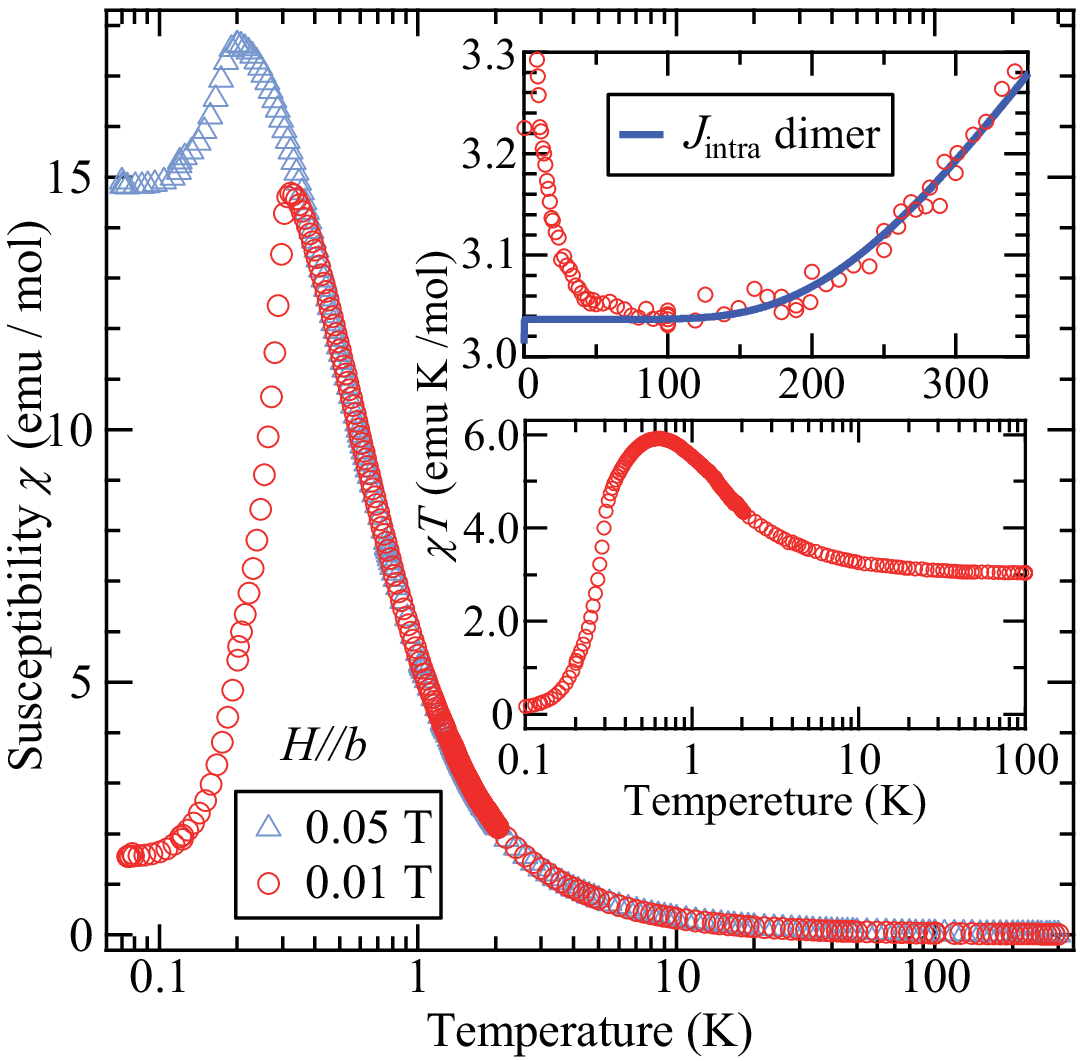}
\caption{(color online) Temperature dependence of magnetic susceptibility ($\chi$ = $M/H$) of [Mn(hfac)$_2$]$\cdot$($o$-Py-V) at 0.01 and 0.05 T for $H//b$. The upper and lower insets show temperature dependence of ${\chi}T$ for high- and low-temperature regions, respectively. The solid line represents the calculated results for the $S_{\rm{Mn}}$-$S_{\rm{V}}$ dimer coupled by $J_{\rm{intra}}$ .}\label{f2}
\end{center}
\end{figure}

Figure 2 shows the temperature dependence of magnetic susceptibility ($\chi$ = $M/H$) for $H//b$.
The temperature dependence of ${\chi}T$ decreases with decreasing temperature and becomes approximately constant around 100 K, as shown in the upper inset of Fig. 2.
This value is very close to the Curie constant of 3.0 emu$\cdot$K/mol for $S$=2 non-interacting spin, which evidences the formation of $S$=2 by the strong AF $J_{\rm{intra}}$. 
We calculated ${\chi}T$ for the $S_{\rm{Mn}}$-$S_{\rm{V}}$ dimer coupled by $J_{\rm{intra}}$ and obtained impressive agreement between the experiment and the calculation above about 100 K by using $J_{\rm{intra}}/k_{\rm{B}}$ = 325 K, where we assumed $g_{\rm{b}}$=2.02 and 99\% of sample purity evaluated from the following ESR and magnetization curve.
Below 100 K, ${\chi}T$ increases with decreasing temperature down to about 0.6 K, as shown in the lower inset of Fig. 2.
This behavior indicates contributions of strong F interaction and/or large anisotropy, which probably arises from a magnetic dipole-dipole interaction.
The Curie-Weiss law, $\chi$ = $C/(T-{\theta}_{\rm{W}})$, describes the system between 5 and 100 K. 
The estimated Curie constant and the Weiss temperature are about $C$ = 3.01 emu$\cdot$ K/mol and ${\theta}_{\rm{W}}$ = +0.78(4) K, respectively.
Below 0.6 K, ${\chi}T$ decreases with decreasing temperature, which indicates the existence of additional weak AF interactions. 
We found distinct sharp peaks in $\chi$ at lower temperatures.
These peaks are associated with phase transitions to a three-dimensional (3D) order because of weak but finite 3D couplings.
In fact, the experimental result of the specific heat $C_{\rm{p}}$ at zero-field exhibits a $\lambda$-type sharp peak associated with the phase transition to the 3D order at $T_{\rm{N}}$ = 0.27 K, as shown in Fig. 3(a).  
Although the lattice contributions are not subtracted from $C_{\rm{p}}$, the magnetic contributions are expected to be dominant in the low-temperature regions considered here.
The finite values at the sufficiently low temperature below $T_{\rm{N}}$ should originate from the contribution of nuclear spins. 
Above $T_{\rm{N}}$, we observed a broad peak at about 0.8 K, which exactly indicates the development of short-range order in the $S$ = 2 F-AF alternating chain. 
For the field direction dependence of $C_{\rm{p}}$, the peak associated with the phase transition for $H//b$ exhibits a significant change, whereas there is no remarkable change for $H//a$ and $H//c$, as shown in Fig. 3(b).
Thus, we can deduce an easy-axis anisotropy along the $b$-axis and consider the on-site anisotropy term in Eq. (1).

Here, we describe the remarkable behavior of the low-temperature magnetization curves for the fields parallel ($H//c$) and perpendicular ($H//b$) to the chain direction.
The measured saturation magnetization $M_{\rm{sat}}\sim $ 4.0 $\mu_{\rm{B}}$/f.u. is consistent with that expected in the present $S$=2 system. 
We observed a distinct metamagnetic phase transition at $H_{\rm{c}}\sim$ 0.06 T for $H//b$, whereas the magnetization curve for $H//c$ apparently exhibits a monotonic increase, as shown in Fig. 4 and its inset.  
This metamagnetic phase transition demonstrates the presence of a large easy-axis anisotropy with the relation of $|D| \gg J_{\rm{AF}}$ if we assume a classical spin system. 
Furthermore, we found an anomalous peak in the field derivative of magnetization ($dM/dH$) for $H//c$.
It should be noted that the value of the magnetization at the peak field of about 0.18 T corresponds to $M_{\rm{sat}}$/4 $\sim$ 1.0 $\mu_{\rm{B}}$/f.u.

\begin{figure}[t]
\begin{center}
\includegraphics[width=16pc]{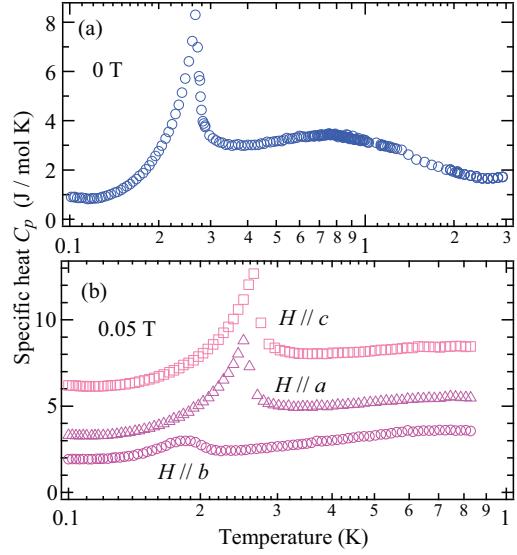}
\caption{(color online) Specific heat of [Mn(hfac)$_2$]$\cdot$($o$-Py-V) at (a) 0 and (b) 0.05 T for the field parallel to the three crystallographic axes. The values for $H//a$ and $H//c$ have been shifted up by 2.0 and 5.0 J/mol K, respectively.}\label{f3}
\end{center}
\end{figure}

\begin{figure}[t]
\begin{center}
\includegraphics[width=18pc]{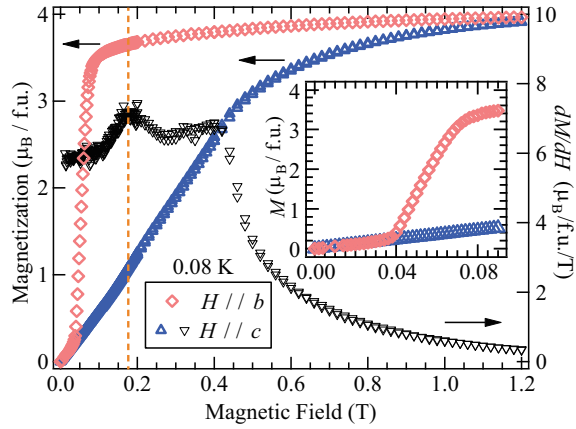}
\caption{(color online) Magnetization curve of [Mn(hfac)$_2$]$\cdot$($o$-Py-V) at 0.08 K for $H//b$ and $H//c$. The vertical broken line shows the field where the $dM/dH$ for $H//c$ exhibits the non-trivial peak. The inset shows the expansion of the low-field region.}\label{f4}
\end{center}
\end{figure}

In order to evaluate the one-site anisotropy, we carried out ESR measurements.
Figure 5(a) shows the temperature dependence of the ESR absorption spectra at 106.6 GHz for $H//b$.
We observed two distinct resonance signals, which are indicated by the large arrows below 10 K. 
In high-temperature regions, the energy states of $S$=2 have a low probability because of the large energy gap between the $S$=2 and $S$=3 states.
With decreasing temperature, the energy states of $S$=2 obtain a higher probability, resulting in the occurrence of the resonance signals originating from the transitions between $S$=2 states.  
Figure 5(b) shows the frequency dependence of the resonance fields at the lowest temperature of 1.5 K, and we plot them in the frequency-field diagram, as shown in the inset.
Considering the energy scale of the interactions and the broad peak temperature of $C_{\rm{p}}$, the magnetic correlation length in the chain direction is expected to be quite short at 1.5 K.
Therefore, we consider the observed signals as $S$=2 paramagnetic resonance and take into account the on-site anisotropy as follows: $\mathcal {H} = D(S_{z})^2+g_{\rm{b}}\mu_{B}HS_{z}$, where the $z$-axis corresponds to the $b$-axis.
The eigenstates defined as $|S_{z}\textgreater$ are split into three levels of $|\pm 2>$, $ |\pm1\textgreater$, and $|0\textgreater$ at zero-field because of the on-site anisotropy, as shown in the inset of Fig. 5(a). 
The application of an external field completely removes such degeneracies.
Accordingly, only transitions from the ground state and the first excited states are mainly observed at low temperatures.
The ESR selection rule permits two resonance modes, $\omega_{1}$ and $\omega_{2}$, which have zero-field energy gaps of 3$D$ and $D$, respectively, as shown in the inset of Fig. 5(a).
The experimental results appear to be associated with these modes.
We obtained $D/k_{\rm{B}}$=$-$0.16(2) K and $g_{\rm{b}}$=2.02(1) through the fitting of the resonance modes, as shown in the inset of Fig. 5(b).

Finally, we discuss the ground state of the present spin model.
Since $H_{\rm{c}}$ depends only on the value of $J_{\rm{AF}}$ in a classical spin system, we evaluated at $J_{\rm{AF}}/k_{\rm{B}}$ = 0.037 K to reproduce the observed magnetization curve for $H//b$ by using the classical Monte Carlo method (see Supplementary). 
Considering the spin Hamiltonian in Eq. (1), the Weiss temperature is given by ${\theta}_{\rm{W}}= -(2D+J_{\rm{F}}+J_{\rm{AF}})S(S+1)/3k_{B}$.
Thus, we can roughly evaluate $J_{\rm{F}}/k_{\rm{B}}$=$-$0.1 K ($|J_{\rm{AF}}/J_{\rm{F}}|$=0.37) by using the above relation and other parameters.
Assuming the Heisenberg case, the strong interaction $J_{\rm{F}}$ in the $S$=2 F-AF chain is expected to form an $S$=4 VBS with an extremely small Haldane gap~\cite{Nakano}. 
In the present case, however, the large anisotropy probably causes a disappearance of such an energy gap and stabilizes a N$\rm{\Acute{e}}$el state.
The anomalous magnetization at $M_{\rm{sat}}$/4 for $H//c$ does not appear in the classical $S$ = 2 F-AF chain with Ising anisotropy (see Supplementary). 
On the other hand, the quantum $S$ = 4 system, which satisfies the necessary condition for realization of a 1/4 magnetization plateau~\cite{oshikawa}, could have a relatively high density of states in the vicinity of the exited state associated with $M_{\rm{sat}}$/4 owing to the discrete nature of the spin.
Therefore, the anomalous magnetization at $M_{\rm{sat}}$/4 should have its origin in the quantum nature of the $S$ = 2 F-AF chain with Ising anisotropy.



\begin{figure}[t]
\begin{center}
\includegraphics[width=17pc]{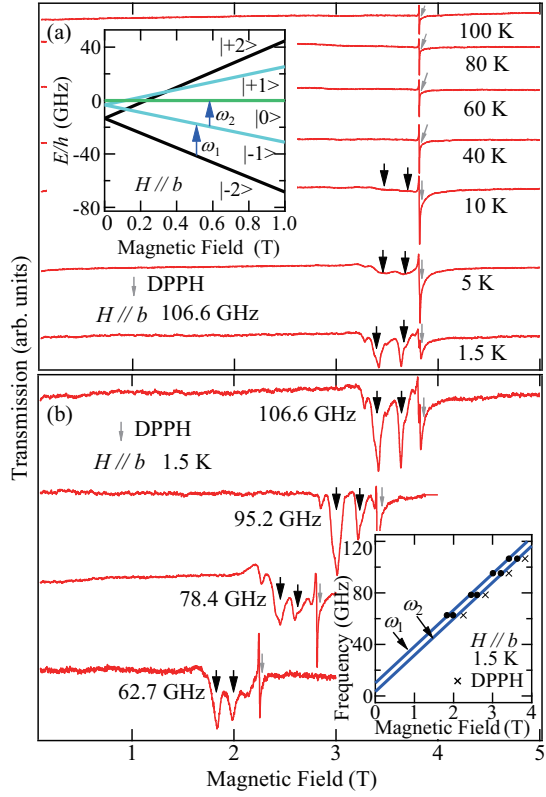}
\caption{(color online) (a) Temperature dependence and (b) frequency dependence of ESR absorption spectra of [Mn(hfac)$_2$]$\cdot$($o$-Py-V) at 106.6 GHz and 1.5 K for $H//b$. DPPH is for correction of the magnetic field. 
The upper inset shows the energy diagram of $S$=2 with $D/k_{\rm{B}}$ = -0.16 for $H//b$, where the arrows indicate probable transitions.
The lower inset shows the frequency-field diagram of the resonance fields. 
The solid lines indicate the probable resonance modes.}\label{f3}
\end{center}
\end{figure}

In summary, we have succeeded in synthesizing a new Mn-verdazyl complex [Mn(hfac)$_2$]$\cdot$($o$-Py-V).
Through the analysis of magnetic susceptibility and the temperature dependence of ESR spectra, we revealed that the strong AF intramolecular interaction between spins on Mn$^{\rm{2+}}$ and the verdazyl radical forms an $S$=2 molecule below about 100 K.
The $ab$ $initio$ MO calculation indicated the formation of the $S$=2 F-AF alternating chain, and we evaluated relatively large on-site anisotropy from the ESR analysis.
These results evidence the first realization of an $S$=2 F-AF alternating chain with Ising anisotropy.  
Furthermore, we observed an anomalous change in the magnetization at 1/4 of the saturation value, which possibly originates from the quantum nature of the present spin model.
Consequently, we demonstrated the realization of $S{\textgreater}$1 quantum spin system through the strong coupling between radical and magnetic ion in the molecular-based complex. 
The present results provide a way to synthesize magnetic materials with a variety of quantum spin
systems with different spin sizes and will stimulate studies on unexplored $S{\textgreater}$1 quantum spin systems.

We thank T. Tonegawa, K. Okamoto, T. Okubo, and T. Shimokawa for the valuable discussions. This research was partly supported by Grant for Basic Science Research Projects from the Sumitomo Foundation, Shorai Foundation for Science and Technology, and  KAKENHI (No. 24540347 and No. 24340075). A part of this work was performed under the interuniversity cooperative research program of the joint-research program of ISSP, the University of Tokyo and the Institute for Molecular Science.



\begin{thebibliography}{99}

\bibitem{Haldane} 
F. D. M. Haldane, Phys. Rev. Lett. \textbf{50}, 1153 (1983).

\bibitem{Todo}
S. Todo and K. Kato, Phys. Rev. Lett. \textbf{87}, 047203 (2001).


\bibitem{Nakano}
H. Nakano and A. Terai, J. Phys. Soc. Jpn. \textbf{78}, 014003 (2009).


\bibitem{NENP}
K. Katsumata, H. Hori, T. Takeuchi, M. Date, A. Yamagishi, J. P. Renard, Phys. Rev. Lett. \textbf{63}, 86 (1989).

\bibitem{CsNiCl3}
I. A. Zaliznyak, S.-H. Lee. and S. V. Petrov, Phys. Rev. Lett. \textbf{87}, 017202 (2001).

\bibitem{Affleck}
I. Affleck, T. Kennedy, E. H. Lieb, and H. Tasaki, Phys. Rev. Lett. \textbf{59}, 799 (1987).


\bibitem{Hagiwara}
M. Hagiwara, K. Katsumata, I. Affleck, B. I. Halperin, J. P. Renard, Phys. Rev. Lett. \textbf{65}, 3181 (1990).


\bibitem{CsCrCl3}
S. Itoh, H. Tanaka, and M. J. Bull, J. Phys. Soc. Jpn. \textbf{71}, 1148 (2002).

\bibitem{MnCl3(bipy)}
G. E. Granroth, M. W. Meisel, M. Chaparala, Th. Jolicoeur, B. H. Ward, and D. R. Talham, Phys. Rev. Lett. \textbf{77}, 1616 (1996).

\bibitem{a-NaMnO2}
C. Stock, L. C. Chapon, O. Adamopoulos, A. Lappas, M. Giot, J. W. Taylor, M. A. Green, C. M. Brown, and P. G. Radaelli, Phys. Rev. Lett. \textbf{103}, 077202 (2009).

\bibitem{MnF(salen)}
T. Birk, K. S. Pedersen, S. Piligkos, C. Aa. Thuesen, H. Weihe, and J. Bendix, Inorg. Chem. \textbf{50}, 5312 (2011).


\bibitem{CrCl2}
M. B. Stone, G. Ehlers, and G. E. Granroth, Phys. Rev. B \textbf{88}, 104413 (2013).


\bibitem{CsCrCl3_2}
S. Itho, T. Yokoo, S. Yano, D. Kawana, H. Tanaka, and Y. Endoh, J. Phys. Soc. Jpn. \textbf{81}, 084706 (2012).



\bibitem{S2_phase1}
H. J. Schulz, Phys. Rev. B \textbf{34}, 6372 (1986).

\bibitem{S2_phase2}
M. Oshikawa, J. Phys. Condens. Matter \textbf{4}, 7469 (1992).

\bibitem{S2_phase3}
H. Aschauer and U. Schollw$\rm{\ddot{o}}$ck, Phys. Rev. B \textbf{58}, 359 (1998).

\bibitem{S2_phase4}
T. Tonegawa, K. Okamoto, H. Nakano, T. Sakai, K. Nomura, and M. Kaburagi, J. Phys. Soc. Jpn. \textbf{80}, 043001 (2011).

\bibitem{S2_phase5}
K. Okamoto, T. Tonegawa, H. Nakano, T. Sakai, K. Nomura, and M. Kaburagi, J. Phys.: Conf. Ser.  \textbf{302}, 012014 (2011).

\bibitem{S2_phase6}
K. Okamoto, T. Tonegawa, T. Sakai, and M. Kaburagi, JPS Conf. Proc. \textbf{3}, 014022 (2014).


\bibitem{S2_phase7}
Y.-C. Tzeng, Phys. Rev. B \textbf{86}, 024403 (2012).


\bibitem{S2_phase8}
J. A. Kj$\rm{\Ddot{a}}$ll, M. P. Zaletel, R. S. K. Mong, J. H. Bardarson, and F. Pollmann, Phys. Rev. B \textbf{87}, 235106 (2013).


\bibitem{S2_alt1}
M. Yamanaka, M. Oshikawa, and S. Miyashita, J. Phys. Soc. Jpn. \textbf{65}, 1562 (1996). 


\bibitem{S2_alt2}
M. Nakamura and S. Todo, Phys. Rev. Lett. \textbf{89}, 077204 (2002).


\bibitem{S2_alt3}
A. Kitazawa and K. Nomura, J. Phys. Soc. Jpn. \textbf{66}, 3379 (1997).


\bibitem{sakai}
T. Sakai, J. Phys. Soc. Jpn. \textbf{64}, 251 (1995).


\bibitem{neut1}
S. Watanabe and H. Yokoyama, J. Phys. Soc. Jpn. \textbf{68}, 2073 (1999).


\bibitem{neut2}
S. Kokado and N. Suzuki, J. Phys. Soc. Jpn. \textbf{68}, 3091 (1999).


\bibitem{hida1}
K. Hida, Phys. Rev. B \textbf{45}, 2207 (1992).

\bibitem{hida2}
K. Hida, J. Phys. Soc. Jpn. \textbf{62}, 1463 (1993).

\bibitem{hida3}
K. Hida, J. Phys. Soc. Jpn. \textbf{67}, 1416 (1998).


\bibitem{3Cl4FV} 
H. Yamaguchi, K. Iwase, T. Ono, T. Shimokawa, H. Nakano, Y. Shimura, N. Kase, S. Kittaka, T. Sakakibara, T. Kawakami, and Y. Hosokoshi, Phys. Rev. Lett. {\bf 110}, 157205 (2013).

\bibitem{mPhV2} 
K. Iwase, H. Yamaguchi, T. Ono, T. Shimokawa, H. Nakano, A. Matsuo, K. Kindo, H. Nojiri, and Y. Hosokoshi, J. Phys. Soc. Jpn. {\bf 82}, 074719 (2013).


\bibitem{26Cl2V} 
H. Yamaguchi, T. Okubo, K. Iwase, T. Ono, Y. Kono, S. Kittaka, T. Sakakibara, A. Matsuo, K. Kindo, and Y. Hosokoshi, Phys. Rev. B {\bf 88}, 174410 (2013).

\bibitem{fine-tune} 
H. Yamaguchi, H. Miyagai, T. Shimokawa, K. Iwase, T. Ono, Y. Kono, N. Kase, K. Araki, S. Kittaka, T. Sakakibara, T. Kawakami, K. Okunishi, and Y. Hosokoshi, J. Phys. Soc. Jpn. {\bf 83}, 033707 (2014).

\bibitem{Zn}
H. Yamaguchi, Y. Shinpuku, T. Shimokawa, K. Iwase, T. Ono, Y. Kono, S. Kittaka, T. Sakakibara, and Y. Hosokoshi, Phys. Rev. B, {\bf 91}, 085117 (2015).


\bibitem{CCDC}
Crystallographic data have been deposited with Cambridge Crystallographic Data Centre: Deposition No. CCDC 1048586 for RT and CCDC 1048588 for 25 K.

\bibitem{MOmetho} 
$Ab$ $initio$ MO calculations were performed using the UB3LYP-D method in the Gaussian 09 program package.
The basis sets are 6-31G++ (O and F) and 6-31G (the other atoms). 
For the estimation of intermolecular magnetic interaction, we applied our evaluation scheme that have been studied previously~\cite{MOcal}. 

\bibitem{MOcal} M. Shoji, K. Koizumi, Y. Kitagawa, T. Kawakami, S. Yamanaka, M. Okumura, and K. Yamaguchi, Chem. Phys. Lett. {\bf 432}, 343 (2006).

\bibitem{oshikawa}
M. Oshikawa, M. Yamanaka, and I. Affleck, Phys. Rev. Lett. {\bf 78}, 1984 (1997).



\end{thebibliography}
\end{document}